\documentclass[12pt]{article}
\usepackage{graphicx}
\usepackage{caption}
\usepackage{subcaption}
\usepackage{lineno}
\newcommand\pubnumber{CMS}
\newcommand\pubdate{\today}

\def\institute{Department of Physics\\
Purdue University}
\def\Title#1{\begin{center} {\Large #1 } \end{center}}
\def\Author#1{\begin{center}{ \sc #1} \end{center}}
\def\Address#1{\begin{center}{ \it #1} \end{center}}

\newcommand\pubblock{\rightline{\begin{tabular}{l} \pubnumber\\
         \pubdate  \end{tabular}}}
\newenvironment{Abstract}{\begin{quotation}  }{\end{quotation}}
\newenvironment{Presented}{\begin{quotation} \begin{center} 
             PRESENTED AT\end{center}\bigskip 
      \begin{center}\begin{large}}{\end{large}\end{center} \end{quotation}}

\def\beq{\begin{equation}}
\def\eeq#1{\label{#1}\end{equation}}
\def\eeqn{\end{equation}}

\def\beqa{\begin{eqnarray}}
\def\eeqa#1{\label{#1}\end{eqnarray}}
\def\eeqan{\end{eqnarray}}

\let\bar=\overbar

\def\Dslash{\not{\hbox{\kern-4pt $D$}}}
\def\dslash{\not{\hbox{\kern-2pt $\del$}}}

\def\msb{{\bar{\ssstyle M \kern -1pt S}}}

\begin{document}
\begin{titlepage}
\pubblock

\vfill
\Title{Top quarks as a probe to quantum information}
\vfill
\Author{Andrew Wildridge on behalf of the CMS Collaboration}
\Address{\institute}
\vfill
\begin{Abstract}
Top quark pairs produced at the Large Hadron Collider (LHC) provide a unique window into quantum information theory at high energies. One of the most ubiquitous measurements of
quantum information is the violation of Bell’s inequality. We explore what would be necessary to observe a violation of Bell’s inequality and the dependence of this on the initial state of the top quark pair.
Furthermore, we show how a more general application of quantum information theory in the realm of quantum computing can be leveraged to perform offline reconstruction of primary vertices. We perform
some optimizations of the running parameters of the quantum annealer and compare to a non-optimized performance. Lastly, we discuss the future outlook of both these topics and steps to be taken.
\end{Abstract}
\vfill
\begin{Presented}
$14^\mathrm{th}$ International Workshop on Top Quark Physics\\
(videoconference), 13--17 September, 2021
\end{Presented}
\vfill
\end{titlepage}
\def\thefootnote{\fnsymbol{footnote}}
\setcounter{footnote}{0}

\section{Introduction}

At the Large Hadron Collider, counter rotating beams containing bunches of protons are passed through each other every twenty-five nanoseconds at designated interaction points. At these interaction points, instrumentation surrounds the bunch crossing such that particle physics events may be observed. The Compact Muon Solenoid (CMS) experiment has instrumentation surrounding one of these such interaction points. At the center-of-mass energies currently accessible by the LHC, the predicted cross section for top-antitop quark pairs is $\sigma_{ttbar} = 831.76^{+19.77}_{-29.20} (scale)^{+35.06}_{-35.06} (PDF+\alpha_S)$ pb \cite{ttbar_crosssection}. Therefore, millions of top-antitop quark pairs will be produced during the Run II operational period of the LHC. This allows for a unique opportunity in performing precision measurements of top quark phenomenon. One such measurement is the measurement of the entanglement and violation of Bell's Inequality in top-antitop quark pairs produced at the LHC.

A different perspective of quantum information being used at the frontier of high energy particle physics is the usage of quantum computing in offline reconstruction tasks. A promising area of research for applying quantum computing in high energy particle physics is the usage of a quantum annealer to perform primary vertexing. Primary vertexing is the clustering of particle tracks to identify the common hard scattering process for these particles. This can be phrased as a combinatorial optimization problem satisfying a single constraint which quantum annealers are well suited to tackling.

\section{Bell's Inequality and Entanglement}
\begin{figure}[!h!tbp]
\centering
\includegraphics[height=1.5in]{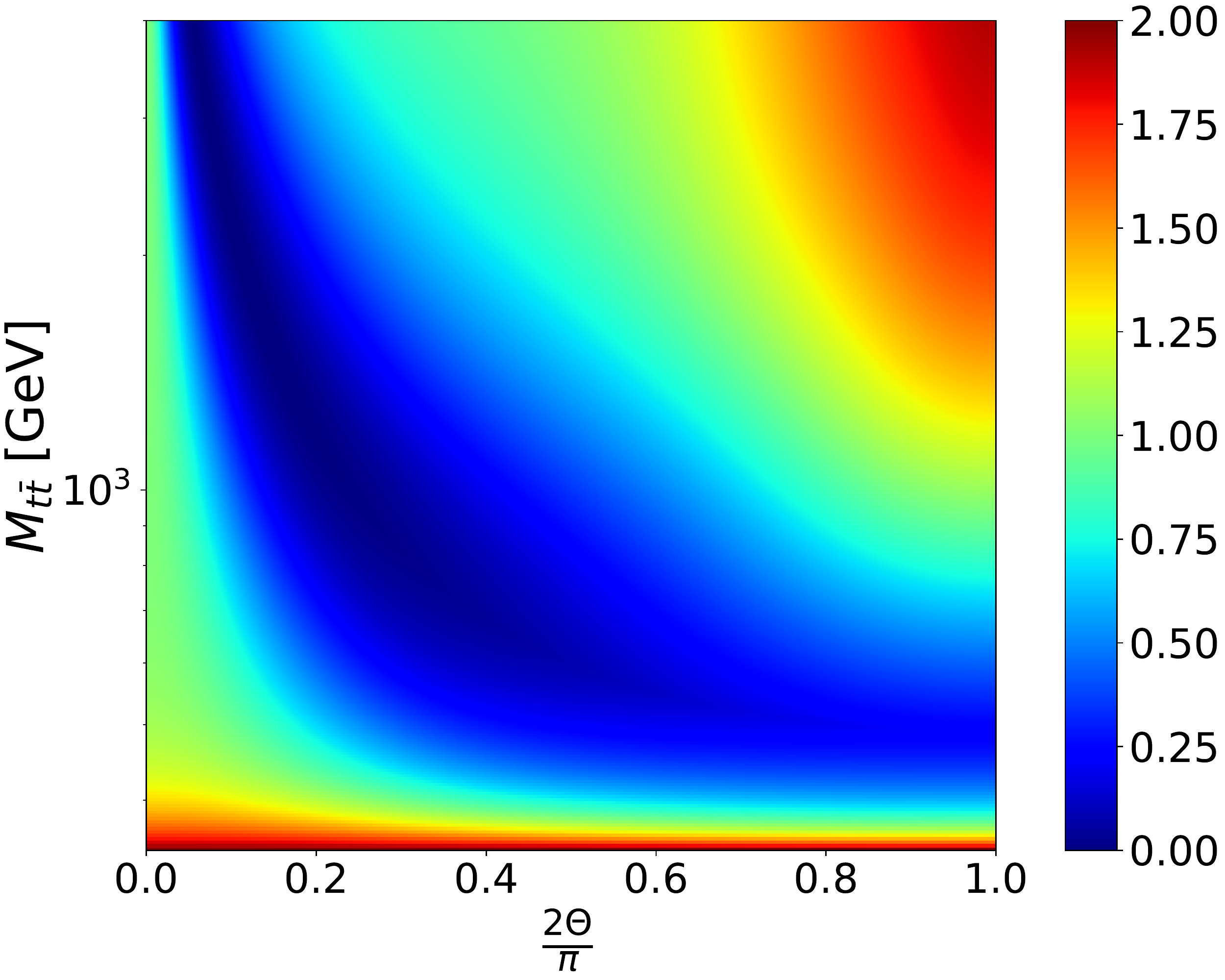}
\caption{The summation of the two largest eigenvalues for the $C^TC$ matrix with initial state gluon-gluon fusion.}
\label{fig:gg_bells}
\end{figure}

\begin{figure}[!h!tbp]
\centering
\includegraphics[height=1.5in]{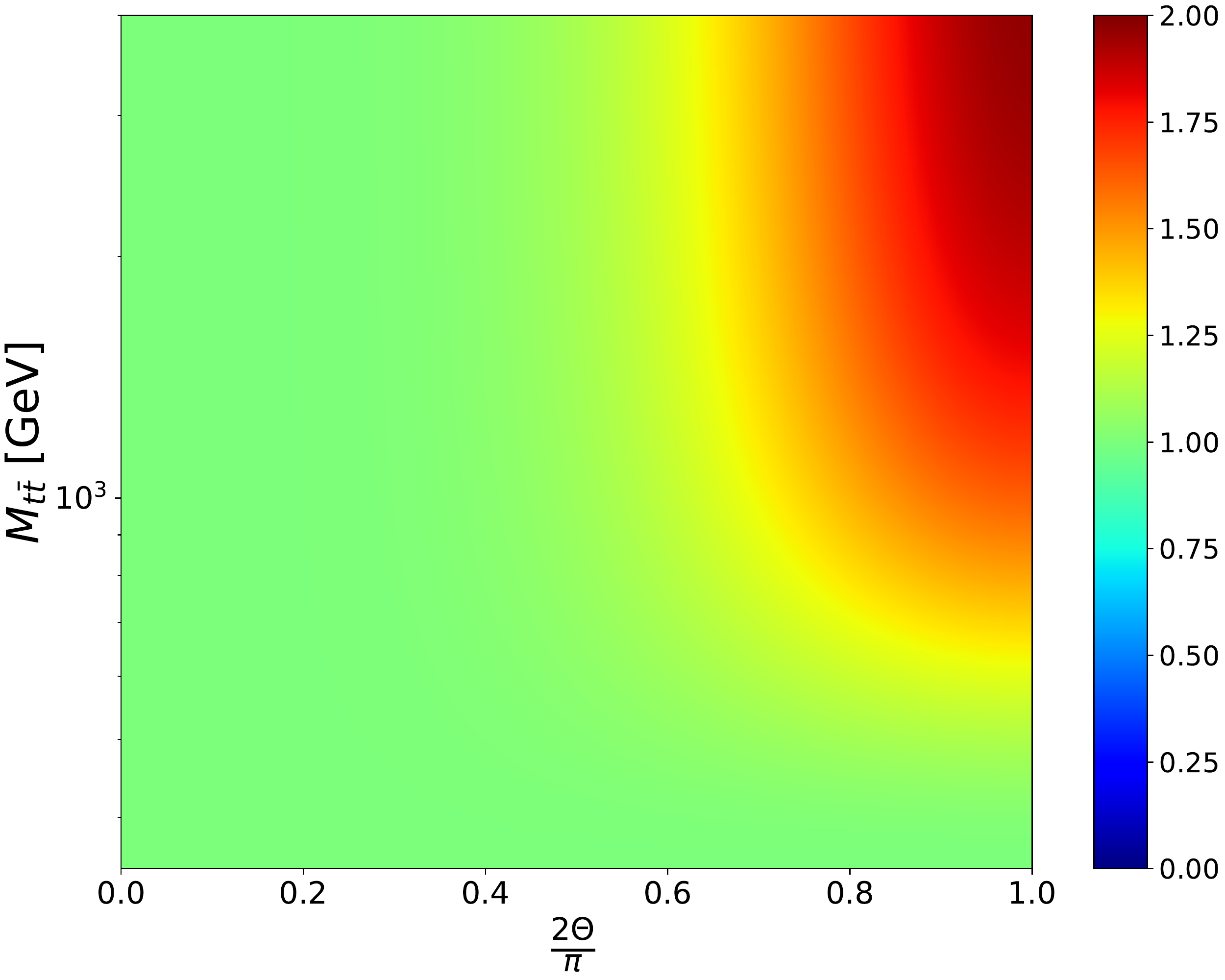}
\caption{The summation of the two largest eigenvalues for the $C^TC$ matrix with initial state quark-antiquark annihilation.}
\label{fig:qq_bells}
\end{figure}

The generalized Bell's inequality for mixed states can be written as follows\cite{bells_ineq}:

\begin{equation}
    |\hat{n}_1 \cdot C \cdot \left( \hat{n}_2 - \hat{n}_4 \right) + \hat{n}_3 \cdot C \cdot \left(\hat{n}_2 + \hat{n}_4 \right)| \leq 2
    \label{generalized_bells_ineq}
\end{equation}

\noindent 
where $\hat{n}_i$, $i \in \{1, 2, 3, 4\}$, are a choice of four different unit vectors along which the spins of the top quark and antitop quark can be measured. Clearly, one must pick these $\hat{n}_i$ such that the left-hand side of Eq.~\ref{generalized_bells_ineq} is maximized. However, by manipulating the left-hand side of Eq.~\ref{generalized_bells_ineq}, one can show that the violation of this inequality is equivalent to the statement\cite{bells_ineq}

\begin{equation}
    m_1 + m_2 > 1
    \label{m1_plus_m2}
\end{equation}

\noindent 
where $m_1$ and $m_2$ are the largest and second largest eigenvalues of the matrix $U = C^TC$. Therefore, the problem of measuring a violation of Bell's inequality for the top-antitop quark pair can be reduced to measuring the spin correlations along some arbitrary basis in some region of phase space. The measurement of the $t\bar{t}$ spin correlations has already been performed using the dilepton final state \cite{spin_corr}. However, the entanglement, and thus the violation of Bell's Inequality, of the $t\bar{t}$ system is highly dependent on three factors: invariant mass of the $t\bar{t}$ system, the scattering angle, and the initial state. The analytic expression for the spin correlation matrix to LO for both the gluon-gluon and quark-antiquark initial states has been derived \cite{entanglement}. Therefore, these expressions can be used to analyze the dependence of the observable in Eq.~\ref{m1_plus_m2} on these three factors. These dependencies are illustrated in Figures \ref{fig:gg_bells} and \ref{fig:qq_bells}.

%%%%%%%%%%%%%%%%%%%%%%%%%%%%%%%%%%%%%%%%%%%%%%%%%%%%%%%%%%%%%%%%%%%%%%%%%
%%
%%   use this format to include an .eps figure into your paper
%%
%%%%%%%%%%%%%%%%%%%%%%%%%%%%%%%%%%%%%%%%%%%%%%%%%%%%%%%%%%%%%%%%%%%%%%%%%%%

By observing Figures \ref{fig:gg_bells} and \ref{fig:qq_bells} it is obvious that any $t\bar{t}$ state produced from an initial $q\bar{q}$ state always violates Bell's inequality. However, the gg fusion production mode is much more difficult to extract signal from and is the dominant production mode for the $t\bar{t}$ state at the LHC with $\sqrt{s} = 13$ TeV. The first obvious region is for large invariant mass of the $t\bar{t}$ system and with a large scattering angle but this region is limited in statistics. The second region is the threshold region where the invariant mass is very close to $2m_t$. This region is notorious for being very hard to reconstruct precisely and with the window of violation of Bell's inequality only sitting at roughly 50 GeV this region of phase space is currently inaccessible.

Entanglement is a necessary condition for violation of Bell's inequality but it is not sufficient\cite{werner}. Therefore, we can expect a measurement of entanglement being present in the $t\bar{t}$ system to be more accessible than a measurement of the violation of Bell's inequality.  By using what is known as the Peres-Horodecki criterion to measure entanglement, it can be shown that the presence of entanglement is equivalent to

\begin{equation}
    \delta \equiv -tr[C] - 1 > 0
    \label{eq:entanglement}
\end{equation}

\noindent 
where C is the spin correlation matrix \cite{entanglement}. Coincidentally, an alternative measurement of the angular separation between the leptons can be performed to extract a related coefficient $D$ given by

\begin{equation}
    D = \frac{tr[C]}{3}
\end{equation}

\noindent 
meaning that a measurement for entanglement present in the $t\bar{t}$ system would simply require an additional phase space cut on D below the critical invariant mass value of ~446 GeV \cite{entanglement} to remove events which are separable. A more elegant cut could also be done to increase the amount of statistics but this is likely unnecessary given the current precision on measuring D \cite{spin_corr} and a significant portion of the cross-section lying below 450 GeV \cite{entanglement}.  

\section{Primary Vertexing with Quantum Annealing}
Primary vertexing can be performed on a quantum annealer by finding the global minimum of the below quadratic unconstrained binary optimization (QUBO) problem

\begin{equation}
    \sum_{i}^{n_T} \sum_{j}^{n_T} \sum_k^{n_V} g\left(D_{ij}; k\right) p_{ik} p_{jk} + \lambda \sum_i^{n_T} \left(1 - \sum_k^{n_V} p_{ik} \right)^2
    \label{eq:qubo}
\end{equation}

\noindent
where $n_T$ is the number of tracks reconstructed in the bunch crossing, $n_V$ is the number of primary vertices requesting to be reconstructed, $p_{ik}$ is the probability that the $i^{th}$ track belongs to the $k^{th}$ primary vertex, $D_{ij}$ is the distance between the reconstructed tracks' location of closest approach to the beam axis, and $g\left(D_{ij}; k\right)$ is a distortion function that modifies the distances to mitigate the effects of noise present in the quantum annealer \cite{primary_vertexing_qa}. The distortion function has a tune-able strength parameter $k$. Each $p_{ik}$ variable is represented by a single logical qubit. For the results presented here the distortion function is

\begin{equation}
    g\left(x; k\right) = 1 - e^{-kx}
    \label{eq:distortion}
\end{equation}

\noindent
with k chosen to be 5 and the distance is

\begin{equation}
    D_{ij} = \frac{|z_i - z_j|}{\sqrt{\delta z_i^2 + \delta z_j^2}}
    \label{eq:distance}
\end{equation}

\noindent
where $z_i$ is the location of closest approach to the beam axis and $\delta z_i$ is the uncertainty in this location.

\begin{figure}[!h!tbp]
\centering
    \begin{subfigure}[b]{0.4\textwidth}
        \centering
        \includegraphics[height=1.5in]{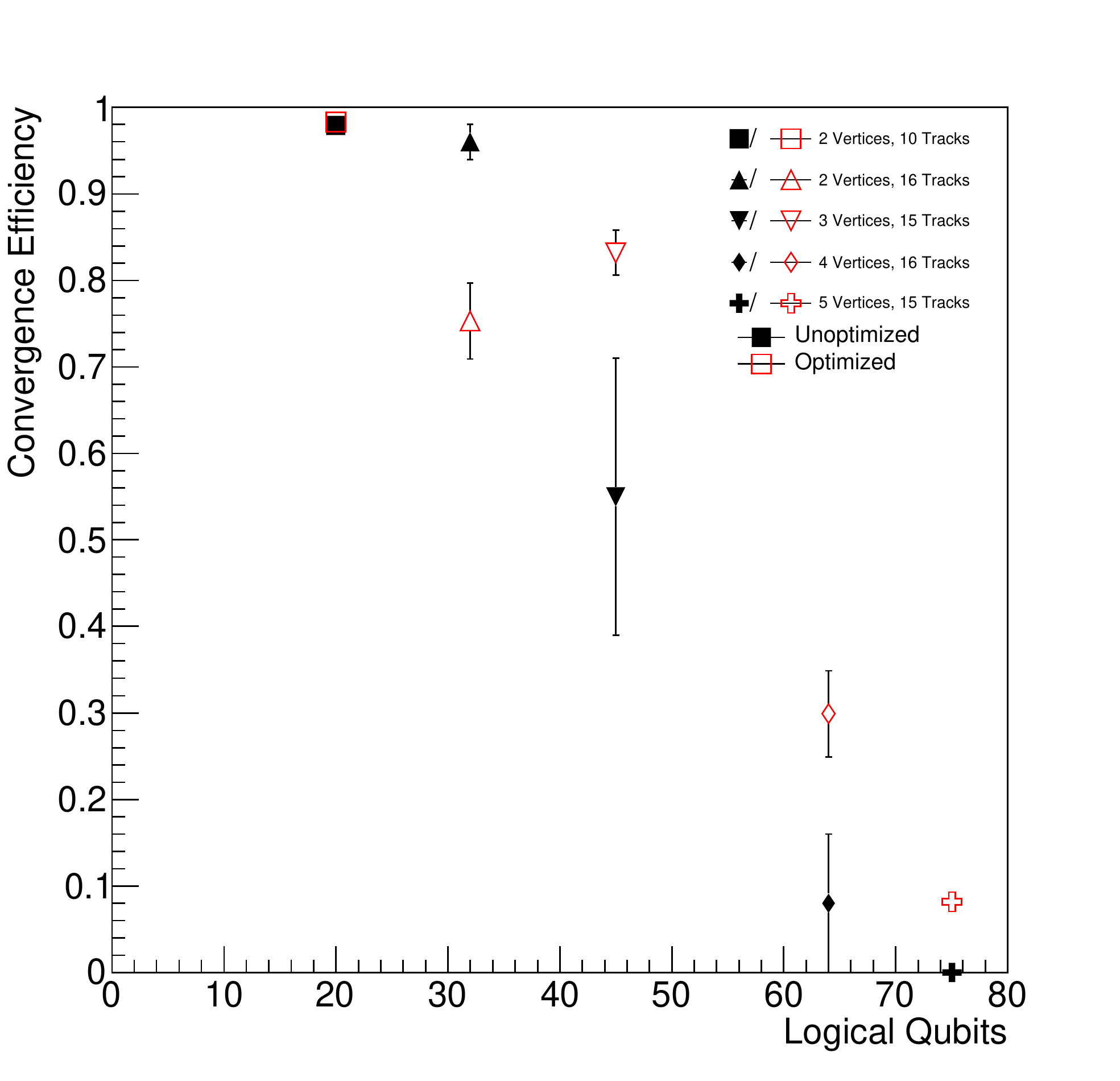}
        \caption{}
        \label{fig:money_plot}
    \end{subfigure}
    \hfill
    \begin{subfigure}[b]{0.4\textwidth}
        \centering
        \includegraphics[height=1.5in]{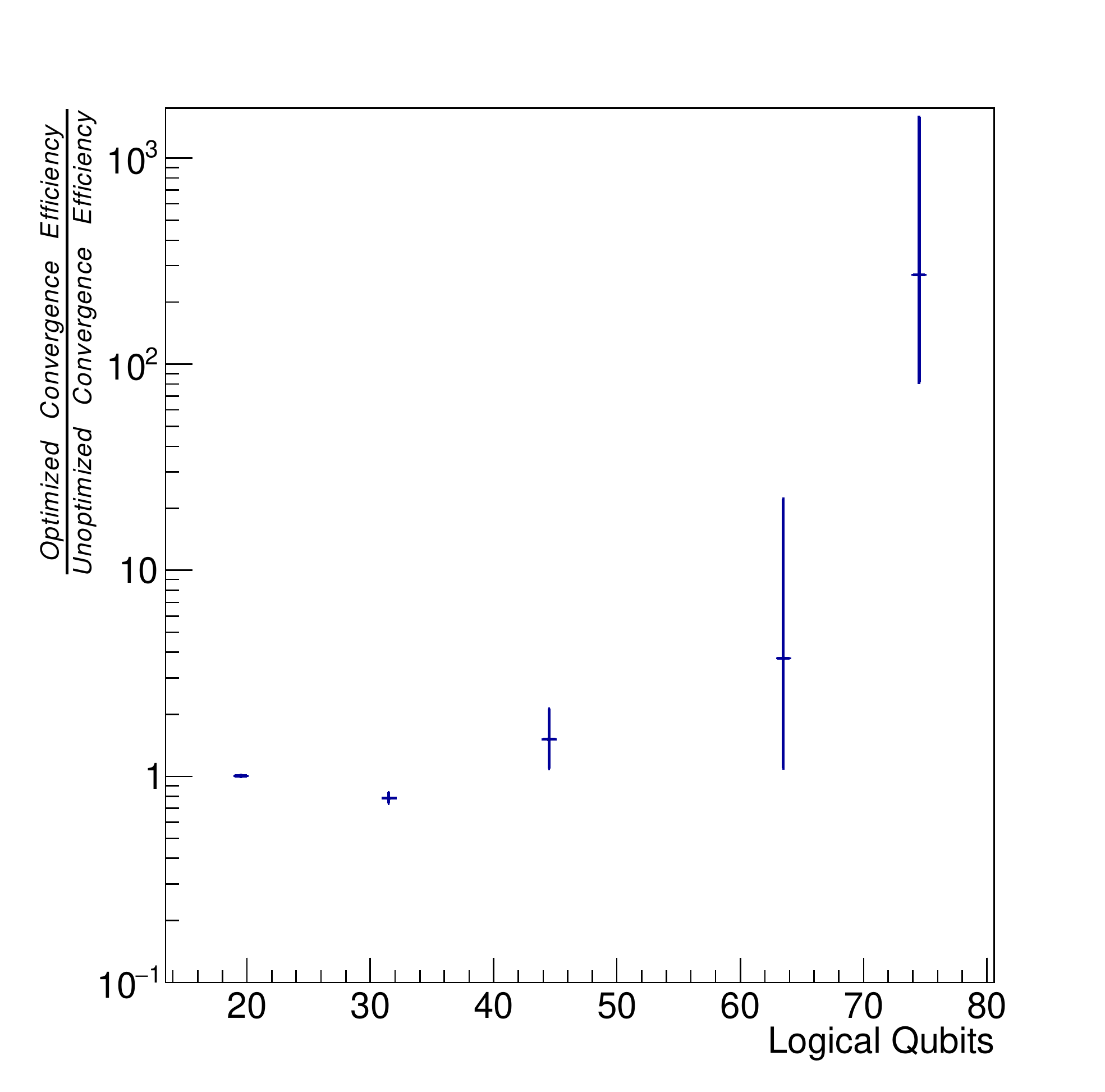}
        \caption{}
        \label{fig:ratio_plots}
    \end{subfigure}
\caption{(Left) Comparison of the optimized quantum annealer's performance versus the default performance for various event topologies. Performance is measured by convergence efficiency which related to the percentage of the time the correct solution was returned.  (Right) Ratio of rate of convergence to the correct solution for the optimized and unoptimized versions of the algorithm.}
\label{fig:optimized_qa}
\end{figure}

The first portion of the QUBO minimizes the cumulative distance between tracks within the same primary vertex. However, the global minimum solution of the first term is to simply put all of the tracks in none of the primary vertices by setting all $p_{ik}$ equal to 0. Therefore, the second term is added as a penalty term to help enforce the constraint that each track must belong to a single primary vertex. Next, multiple running parameters of the quantum annealer were optimized to fully leverage the capabilities of the quantum annealer. This included programming a deterministic embedding of our problem onto the DW 2000Q annealer, applying an optimal chain strength dependent on the lengths of the chains in the embedding, and the overall amount of time given to the annealer during the anneal to minimize the Time-To-Solution (TTS) metric. After these running parameters were optimized, a very large improvement in the performance of the quantum annealer was recorded and can be observed in Figure~\ref{fig:optimized_qa}. Various event topologies were studied with the greatest performance boost being observed for the largest event topology of 5 primary vertices and 20 particle tracks.

\section{Outlook}

A measurement of the violation of Bell's inequality in the $t\bar{t}$ system produced at the LHC presents itself as a challenging multidimensional differential measurement. Many advances will likely be required to observe a violation of Bell's inequality such as an initial state classifier, improved reconstruction resolution, or the full Run III dataset. However, a measurement of whether there is entanglement present in the $t\bar{t}$ system seems well within the grasp using current methods and data from the full Run II dataset. Both measurements are exciting prospects at the frontier of precision physics in the top quark sector.

Primary vertexing on a quantum annealer is currently out-of-range for the event topologies typically encountered with the LHC and are surely out-of-range for the HL-LHC upgrade. However, it is noteworthy that this primary vertexing algorithm could have been used for Tevatron experiments. With the exponential growth happening in quantum annealers, it remains to be seen whether quantum annealers will start to outpace the demands set forth for primary vertexing in the near future. As it stands currently, performing optimizations of the running parameters of a quantum annealer allows for many-fold improvements in algorithm convergence to the correct solution. With a slew of running parameters to choose from to optimize, it is still unclear which parameters are the most important to optimize. Therefore, further optimizations may allow for an even greater increase in algorithm performance such as pausing during the anneal, reverse annealing, and anneal offsets.

%%%%%%%%%%%%%%%%%%%%%%%%%%%%%%%%%%%%%%%%%%%%%%%%%%%%%%%%%%%%%%%%%%%%%%%%%
%%
%%   use this format to include a LaTeX table  into your paper
%%
%\begin{table}[!h!tbp]
%\begin{center}
%\begin{tabular}{l|ccc}  
%col 1 &  col 2 &  col 3 [GeV]&  
%col 4\\ \hline
% row 1  &   0.12     &     10      &     0.1  \\
% row 2 &   0.15     &     100     &  $\pm 10$ \\ \hline
%\end{tabular}
%\caption{Brief table caption.}
%\label{tab:ex}
%\end{center}
%\end{table}
%%%%%%%%%%%%%%%%%%%%%%%%%%%%%%%%%%%%%%%%%%%%%%%%%%%%%%%%%%%%%%%%%%%%%%%%%%%

\bibliography{eprint}{}
\bibliographystyle{unsrt}
 
\end{document}